\documentclass[draft]{article}

\newcommand{\abs}[1]{\ensuremath{\vert #1 \vert}}

\newcommand{\feynmanslash}[1]{\ensuremath{#1\kern -0.5em / }}
\newcommand{\diag}{{\rm diag}}
\newcommand{\us}{\ensuremath{\hat{{\bf s}}}}
\newcommand{\unitvec}[1]{\ensuremath{\hat{{\bf #1}}}}
\newcommand{\ket}[1]{\ensuremath{\vert #1 \rangle}}
\newcommand{\bra}[1]{\ensuremath{\langle #1 \vert}}

\newcommand{\ie}{{\it i.e.}}

\title{Berry phase for spin--1/2 particles moving
            in a spacetime with torsion}
\author{M. Alimohammadi\thanks{e-mail:
           {\tt alimohmd@theory.ipm.ac.ir}} $^1$
           ~and A. Shariati\thanks{e-mail:
           {\tt shahram@iasbs.ac.ir}} $^{2,3}$
 \\[0pt]
  \\[0pt] $^1$ {\it\small Department of Physics, Faculty of Sciences, Tehran
  University,} \\[0pt] {\it\small North Karegar Ave., Tehran, Iran.}
  \\[0pt] $^2$ {\it\small Institute for Advanced Studies in Basic Sciences,}
  \\[0pt] {\it\small P.O. Box 159, Zanjan 45195, Iran.}\\
  $^3${\it\small Institute for Studies in Theoretical Physics
  and Mathematics,}
 \\[0pt] {\it\small P.O. Box 5531, Tehran 19395, Iran.}
} 

\begin{document}
\maketitle

\begin{abstract}
Berry phase for a spin--1/2 particle moving in a flat spacetime with
torsion is investigated in the context of the Einstein-Cartan-Dirac model.
It is shown that if the torsion is due to a
dense polarized background, then there is a Berry phase only if the fermion
is massless and its momentum is perpendicular to the direction of the
background polarization. The order of magnitude of this Berry phase is discussed
in other theoretical frameworks.
\end{abstract}
\begin{center}
{\bf PACS :} 04.50.+h, 03.65.Bz
\end{center}
\newpage

\section{Introduction}
The geometry of a four dimensional spacetime $U^4$ is given by the metric
($g_{\mu\nu}$) and torsion ($T^{\alpha}_{{}\mu\nu}$) tensors.
In the context of Einstein-Cartan-Dirac (ECD) theory,
the axial current of the background material is the source of the
torsion field and leads to a completely anti--symmetric torsion tensor,
represented by a pseudo--vector
$S^{\mu}$. This pseudo--vector is coupled to the axial
current of all the fermion species.
(For a brief review of Einstein-Cartan-Dirac theory,
see appendix \ref{ECD-review}.)

One of the most important features of the physics of torsion
is its phenomenological aspects.
To study this, we must realize that in the context of ECD theory, the torsion
of space--time vanishes in vacuum, and one expects a
non--vanishing torsion pseudo--vector
only if the space is filled
with a (spin--) polarized background matter.
But in such a space--time,
there are plenty of interactions which can easily mask any effect
of torsion. So the best candidate to probe such an interaction is
neutrino, as it is weakly coupled to the rest of matter.

Studies on the interactions of neutrinos with torsion, go to
several years ago. In \cite{dg} (see also \cite{ds}), the effect
of torsion on neutrino oscillation has been studied by assuming
that the torsion eigenstates, \ie\ the eigenstates of the interaction
part of the Hamiltonian, are different from the weak interaction
eigenstates. Hammond has studied the different aspects of fermions'
interaction with a torsion field derived from a second rank potential
\cite{h1,h3}, for example the torsion coupling constant
\cite{h4}
and the relation between the intrinsic spin of the string and the torsion
\cite{h5}. The theoretical and phenomenological aspects of torsion field have
been investigated in \cite{bs} by an effective field theoretical method, and
the contribution of the torsion of
space--time on standard neutrino oscillation has been studied in the context of
ECD theory in \cite{as1}, in which the torsion and weak interaction
eigenstates has been considered the same.
More recently, the quantum reflection of a
massless neutrino from a torsion induced potential barrier has been discussed
in \cite{as2}.

Before going further, it may be useful to describe why we consider the ECD
theory to investigate the physics of the torsion and why we do not work in
a more general framework in which the torsion field is considered as an
propagating quantum field. The reason, in our view, is that if one considers the
torsion field $S^\mu$ as a quantum field which propagates, the resulting theory
will have serious problems. As has been shown in \cite{ber}, the effective
quantum field theory of a massive fermion coupled to the axial vector $S^\mu$
(i.e. the torsion field) is unitary and renormalizable only when $m<<M$; that
is when the torsion mass is much greater than the mass of the heaviest fermion.
But the restrictions coming from the contact experiment achieve only the region
$M<3$Tev \cite{bs}, which is not enough to satisfy $m<<M$ condition for all the
fermions of the standard model. Therefore the ECD theory is almost the unique
avialable quantum theory of gravity with torsion.

In general, the evolution of the neutrinos in a space--time is affected by:
1) the structure of the mass matrix (normally leading to oscillations);
2) effects of matter ( weak interaction, MSW effect, etc.);
3) gravity (\ie\ metric); and
4) torsion of space--time.
In general, both the amplitude and the phase of the neutrinos' wavefunctions
can be affected by all these effects.
Now if there is some nontrivial contribution to the phase of neutrino's
wavefunction due to the
torsion field---the so called Berry phase---then there might be some detectable
effect associated with torsion. So it is worthy to investigate the Berry phase
of a spin--$1/2$ particle in a space--time with torsion.

In this paper, we want to study this Berry phase in the
context of ECD theory.
We consider a spacetime $U^4$, whose metric is
$\eta_{\mu\nu} = \diag (1, -1, -1, -1)$, and its torsion pseudo--vector
$S^{\mu}$ is due to a static polarized dense matter.%
\footnote{Actually, for torsion to be important, the matter
          must be very dense; and in such a situation the
          metric shall not be flat. We are considering a flat
          metric to see the effect of a pure torsion.
          Any calculation must be repeated in the more
          general case of a curved metric, such as the
          Schwarzschild metric.} 
We see that $S^{\mu} = (0, K'\us)$, where $K'$
is some constant depending on the specific model considered---through the coupling
constant---and also on the density and polarization of the background matter.
The unit vector \us\ determines the direction of the background polarization (spin) .

\section{Dirac equation in $U^4$}
The Hamiltonian for a spin--1/2 particle moving in this $U^4$ spacetime
is $H = H_0 + H_1$ (see eq.(\ref{20})), where $H_0 =c{\bf \alpha}\cdot{\bf P}+mc^2 \beta$
is the usual Hamiltonian in a flat (Minkowski) spacetime
($\alpha_1$, $\alpha_2$, $\alpha_3$, and $\beta$ are
Dirac matrices),
and $H_1 =(\hbar c/8)\gamma_5\gamma^0\feynmanslash{S}$
is the interaction Hamiltonian due the torsion field.

It is important to note that both $H_{0}$ and $H_{1}$ are
Hermitian. Therefore, the time evolution generated by the
total Hamiltonian is unitary. So the conclusion of the
articles \cite{cls} and \cite{cils} where the authors have obtained a
``dissipative term'', which causes the state to decrease
exponentially, is wrong. Their mistake, we think,
is in calculating $H_{1}$ (eq.(10) of \cite{cils}).
More recently, the effect of Berry phase on neutrino oscillation
has been studied in \cite{cl}. In these papers, the
authors have considered a null vector $S^\mu$, which is different from ours
(which is derived in the framework of ECD theory).
Also they have considered the massive neutrinos, which again is different from
our situation. As we will show,
we have to consider the
massless neutrinos ( which can not oscillate).

In order to study the Berry phase of a spin--1/2 particle, we must first
calculate the torsion field $S^\mu$. For simplicity, we consider a fermionic
medium  with all the fermions {\it at rest}, through which our spin--1/2 particle moves.
In this case, it can be shown that the torsion field $S^\mu$ is
$S^\mu =(0,K'\us)$, where $K'$ in the context of ECD theory is $-48\pi\hbar\rho G/c^3$, and
$\rho$ and $\us$ are the number density and polarization unit vector of the
background matter, respectively \cite{as2}.
Now choosing the chiral representation
for the Dirac matrices, these two Hamiltonians read
\begin{equation}
H_0 = \pmatrix{-c{\bf \sigma}\cdot{\bf P} & mc^2 \cr mc^2 & c{\bf \sigma}\cdot{\bf P}},
\; H_1 = K\pmatrix{{\bf \sigma}\cdot\us & 0 \cr 0 & {\bf \sigma}\cdot\us},
\end{equation}
where the coupling constant $K$ is (in ECD model)
\begin{equation}\label{couple}
K_{\rm{ECD}}={12\pi\rho G \hbar^2 \over c^2}.
\end{equation}
Here ${\bf P}$ is the momentum of the spin--1/2 particle, and $\sigma_i$s are
Pauli matrices.
Let's take ${\bf P} = P\, \unitvec{z}$ and
$\us = \sin\varphi \cos\alpha \, \unitvec{x} + \sin\varphi \sin\alpha\,
\unitvec{y} + \cos\varphi\, \unitvec{z}$. Therefore the total Hamiltonian becomes
\begin{equation}
H = \pmatrix{-cP + K s_{3} & K s_{-} & mc^2 & 0 \cr K s_{+} & cP - K S_{3} & 0 & mc^2
\cr mc^2 & 0 & cP + K s_{3} & K s_{-} \cr 0 & mc^2 & K s_{+} & -cP-K s_{3} },
\end{equation}
where $s_{\pm}:=s_{1}\pm is_{2}$ and $s_{3}:=s_{z}$.

\section{Evolution of particle's state and its Berry phase}
We consider the following problem: An eigenstate of $H_0$ begins to move in
$U^4$. This can be the case if, for example, there is a region of space
where the matter is polarized and a spin--1/2 particle enters this region;
or if in such a region, a particle is created as an eigenstate of $H_0$.

With no loose of generality, we can Choose \us\ in the $xz$ plane.
In this case, the eigenvalues of $H$ are $E_1$, $-E_1$, $E_2$, and $-E_2$, where
\begin{eqnarray}
E_1  =  \sqrt{K^2 + c^2P^2 + m^2c^4 + 2c K \sqrt{m^2c^2+P^2\cos^2\varphi}}, \\[5pt]
E_2  =  \sqrt{K^2 + c^2P^2 + m^2c^4 - 2c K \sqrt{m^2c^2+P^2\cos^2\varphi}}.
\end{eqnarray}
Let \ket{\psi(0)} be the following eigenstate of $H_0$
\begin{equation} \label{psi0}
\ket{\psi(0)} = \pmatrix{0 & 1 & 0 & (\sqrt{q^2+m^2c^2}-q)/mc}^{{\rm t}},
\end{equation}
where $q$ is the momentum of the particle in torsion--free region and "t"
denotes transpose.
For $m=0$, this state becomes $\ket{\psi(0)} = \pmatrix{0&1&0&0}^{{\rm t}}$,
which is the spinor of a left--handed neutrino.
\ket{\psi(0)} in eq.(\ref{psi0}) can be written as a superposition of \ket{E_1},
\ket{-E_1}, \ket{E_2}, and \ket{-E_2}:
\begin{equation} \label{psi0expand}
\ket{\psi(0)} = a\, \ket{E_1} + b\, \ket{-E_1} + c\, \ket{E_2}+ d\, \ket{-E_2}.
\end{equation}
At time $t$, this state becomes
\begin{equation}
\ket{\psi(t)} = a\,e^{-i E_1 t/\hbar}\, \ket{E_1}+b\, e^{i E_1 t/\hbar}\, \ket{-E_1}
+c\, e^{-i E_2 t/\hbar}\, \ket{E_2}+d\, e^{i E_2 t/\hbar}\, \ket{-E_2}.
\end{equation}
From the general theory of Berry phase \cite{aa}, we know that if at some
time $t$, say $T$, $\ket{\psi(T)} = e^{i\Phi}\, \ket{\psi(0)}$, then
there will be a Berry phase
$ \beta = \Phi + i\int_0^T {\rm d}t\, \bra{\psi(t)}{\rm d}/{\rm d}t
\ket{\psi(t)}.$ In our problem, such a $T$ exists only if
$\exp(2iE_1 T/\hbar) = \exp(i(E_1-E_2)T/\hbar) = \exp(i(E_1+E_2)T/\hbar) =1,$
from which it follows that $E_2 = (2n+1) E_1$, for some integer $n$.
In other words,
\begin{equation}
4n(n+1) (K^2 +c^2 P^2 + m^2c^4) + 4 (2n^2 + 2n +1)c K \sqrt{m^2c^2 + P^2 \cos^2\varphi}
= 0.
\end{equation}
This is independent of $P$, only if $n=0$. We then get
$K \sqrt{m^2c^2 + P^2 \cos^2\varphi}= 0$, which is true in either of the following
two cases:
\begin{enumerate}
\item $K =0$, \ie, when there is no torsion. This is the case of a free
particle moving in ordinary, \ie\ torsion free, Minkowski
spacetime. In this case there is no Berry phase, the phase is only dynamical.
\item $m^2c^2 + P^2 \cos^2\varphi =0$, which is true only if $m=0$ and
$\varphi = \pi/2$. In the following, we calculate the Berry phase for
this nontrivial case, \ie\ a massless fermion with momentum perpendicular
to the polarization of the background.
\end{enumerate}
To calculate the Berry phase in case 2 above, we need the eigenstates
\ket{E_1}, \ket{-E_1}, \ket{E_2}, and \ket{-E_2} (when $m=0$ and
$\varphi = \pi/2$, we have $E_1=E_2=E$)
\begin{eqnarray}
\ket{E_1} =\pmatrix{
                     \frac{-cP+E}{\sqrt{K^2+(cP-E)^2}} \cr
                     \frac{K}{\sqrt{K^2+(cP-E)^2}}    \cr
                     0                               \cr
                     0
                    } 
,
\ket{-E_1} = \pmatrix{
                      0                              \cr
                      0                              \cr
                      \frac{cP-E}{\sqrt{K^2+(cP-E)^2}} \cr
                      \frac{K}{\sqrt{K^2+(cP-E)^2}}
                      }
,
\end{eqnarray}
\begin{eqnarray}
\ket{E_2} = \pmatrix{
                     0                              \cr
                     0                              \cr
                     \frac{cP+E}{\sqrt{K^2+(cP+E)^2}} \cr
                     \frac{K}{\sqrt{K^2 + (cP+E)^2}}
                      }
,
\ket{-E_2} = \pmatrix{
                      -\frac{cP+E}{\sqrt{K^2 + (cP+E)^2}} \cr
                       \frac{K}{\sqrt{K^2 + (cP+E)^2}}   \cr
                       0                                \cr
                       0
                      }
,
\end{eqnarray}
where $E := \sqrt{c^2P^2 + K^2}$ is the energy of \ket{\psi(0)}.

Expanding (\ref{psi0}) (with $m=0$) in terms of these
spinors, and using the notation of (\ref{psi0expand}), we see that
$b=c=0$, and
\begin{equation}
i\bra{\psi(t)}{\rm d}/{\rm d}t \ket{\psi(t)} = {E\over \hbar} (\abs{a}^2 - \abs{d}^2)
={1\over \hbar}\sqrt{E^2-K^2}.
\end{equation}
Also in this case, $\Phi$ and $T$ are
\begin{equation}
\Phi =-\pi , \; T={\pi \hbar \over E}.
\end{equation}
Therefore, the Berry phase $\beta$ becomes
\begin{equation}
\beta =\pi \left( \sqrt{1-(K/E)^2}-1 \right)=-{\pi \over 2}{K^2\over E^2}+\cdots .
\end{equation}
This means that $\lim_{K\to 0} \beta = 0 $, as we expect.
Note that $K$ has dimension of energy.

The order of magnitude of this effect depends on the
coupling constant and the total axial current of the background
matter. In Einstein-Cartan-Dirac theory (minimal coupling),
$K$ is given by (\ref{couple}) which leads to the small value
$K_{\rm{ECD}} ({\rm eV}) \sim 10^{-69}\rho\, ({\rm cm}^{-3})$.
But in other theoretical frameworks, it may lead to greater values.
For example in some models, the Newton's
gravitational constant is replaced by the weak coupling constant
$G_{T} \sim 10^{31} G$ (see for example \cite{ds}), which leads to
$K_{\rm{V-A}} ({\rm eV}) \sim 10^{-38}\rho\, ({\rm cm}^{-3})$.
In the effective field theory approach of Belyaev and Shapiro \cite{bs},
the value of $K$ could be as big as
$K_{\rm{EFT}} ({\rm eV}) \sim 10^{-38}\rho\, ({\rm cm}^{-3})$.
And Finally in strong gravity regime (\ie\ inside a collapsing matter
or in the early stage of the universe), $G\rightarrow G_{\rm{SG}}\sim 10^{39}G$
\cite{siv}, so
$K_{\rm{SG}} ({\rm eV}) \sim 10^{-30}\rho\, ({\rm cm}^{-3})$. For a more detailed
discussion on the torsion coupling constant, see \cite{as2} and references therein.

\appendix \section{Brief review of Einstein-Cartan-Dirac theory}
\label{ECD-review}
The geometry of a $d$--dimensional spacetime $U^d$ is given by two
geometrical objects: a metric tensor $g_{\mu\nu}$
and a connection $\Gamma^{\alpha}_{{}\mu\nu}$ .
The most general metric--compatible
connection is
$\Gamma^{\alpha}_{{}\mu\nu} = \left\{^{\alpha}_{{}\mu\nu}\right\} + %
K^{\alpha}_{{}\mu\nu}$, where $\left\{^{\alpha}_{{}\mu\nu}\right\}$ is
the usual Christofell symbol, derived from the metric, and
$K^{\alpha}_{{}\mu\nu}$ is a tensor of rank 3, named contorsion.
$K^{\alpha}_{{}\mu\nu}$ is related to torsion tensor
$T^{\alpha}_{{}\mu\nu}:=\Gamma^{\alpha}_{{}\mu\nu}-\Gamma^{\alpha}_{{}\nu\mu}$
as follows
\begin{equation}
K_{\alpha\mu\nu}={1\over 2}(T_{\alpha\mu\nu}-T_{\mu\alpha\nu}-T_{\nu\alpha\mu}).
\end{equation}
$T_{\alpha\mu\nu}$ can be decomposed as
\begin{equation} \label{decomposition}
T_{\alpha\mu\nu} = \frac{1}{3}
\left( g_{\alpha\nu} T_{\mu} -  g_{\alpha\mu} T_{\nu} \right)
- \frac{1}{6}  \varepsilon_{\alpha\mu\nu\sigma}S^{\sigma} +
q_{\alpha\mu\nu},
\end{equation}
where $T_{\mu} :=- g^{\alpha\beta} K_{\alpha\beta\mu}$,
$S^{\sigma} := -\varepsilon^{\sigma\alpha\mu\nu} K_{\alpha\mu\nu}$,
and $q_{\alpha\mu\nu}$ is the reminder, defined by equation
(\ref{decomposition}).
Using the usual procedure,
one can show that the scalar curvature $R$ of this space--time is
\begin{equation}
R = \tilde{R} - \frac{2}{\sqrt{-g}}\partial_{\kappa}\left(\sqrt{-g}\,
\tau^{\kappa}\right) - \left( \frac{4}{3} T^2+
\frac{1}{24} S^2+\frac{1}{2} q_{\alpha\mu\nu}\, q^{\mu\nu\alpha} \right),
\end{equation}
where $\tilde{R}$ is the Ricci scalar derived from the Christofell
symbols, \ie\ the scalar curvature of the torsion free space--time.

In Einstein-Cartan-Dirac theory, the spacetime is assumed to be a $U^4$, with
both metric and torsion. The fields of the theory are: the metric, the
contorsion, and a set of spin--$1/2$ fields which are minimally coupled
to the metric and torsion by the usual covariant derivative. The total
action of Einstein-Cartan-Dirac theory is $I = I_{{\rm EC}} +
I_{{\rm D}}$, where
\begin{equation}
I_{{\rm EC}} = -\frac{c^3}{16\pi G} \int d^4 x \sqrt{-g}\, R,
\end{equation}
and
\begin{equation}
I_{{\rm D}} = i\hbar \sum_j\int d^4 x \sqrt{-g}\,
\bar{\psi_j} \left( e^{\mu}_{a} \gamma^{a} (\partial_{\mu} + \Gamma_{\mu})
+i \frac{m_jc}{\hbar} \right) \psi_j,
\end{equation}
where the sum is over all fermions species.
Variation with respect to the contorsion field leads to
\begin{equation}\label{A}
S^{\mu} =  \frac{72\pi\hbar G}{c^3} \sum_j(J_j)_{5}^{\mu},
\end{equation}
\begin{equation}
\; T^{\mu} =0, \;
q_{\alpha\mu\nu} = 0.
\end{equation}
This means that, the contorsion (or torsion) field is completely
anti--symmetric, and the pseudo--vector dual to the torsion field
is the sum of the axial currents of the
fermion field(s)
($(J_j)_{5}^{\mu}=\bar{\psi_j}\gamma^\mu
\gamma_5\psi_j$ ).

Variation with respect to the fermion fields, leads to the following
Dirac equation
\begin{equation}
\gamma^\mu\partial_\mu\psi_j+i\frac{m_j c}{\hbar}\psi_j
+\frac{i}{8}\gamma_5\feynmanslash{S}\psi_j=0.
\end{equation}
This equation can be written as a Schr\"{o}dinger--type equation,
$i\hbar\partial_{t} \psi = H \psi $,
where for $g_{\mu\nu}=\eta_{\mu\nu}$,
the Hamiltonian $H$ becomes
\begin{equation}\label{20}
H = c {\bf\alpha}\cdot{\bf P} + m c^2 \beta
+\frac{\hbar c}{8} \gamma_5\gamma^0\feynmanslash{S}.
\end{equation}
In the above equations, $S^\mu$ is the torsion pseudo--vector
where in the Einstein-Cartan-Dirac
theory is given by eq.(\ref{A}).

\vspace {1cm}

\noindent{\bf Acknowledgement}

\noindent M. Alimohammadi would like to thank the Institute for Studies in
Theoretical Physics and Mathematics and also the research council of the
University of Tehran, for their partial financial supports.

\end{document}